%% file: main.tex
\documentclass[%
 reprint,
 amsmath,amssymb,
 aps,
]{revtex4-1}

\usepackage{graphicx}% Include figure files
\usepackage{dcolumn}% Align table columns on decimal point
\usepackage{bm}% bold math
\usepackage{braket}
\usepackage{color}
\newcommand{\average}[1]{\ensuremath{\langle#1\rangle}}
\begin{document}

\preprint{APS/123-QED}

\title{Quantum Critical Phenomena in Heat Transport via a Two-State System}% Force line breaks with \\

\author{Tsuyoshi Yamamoto}
\email{t.yamamoto@issp.u-tokyo.ac.jp}
\author{Takeo Kato}%
% \email{kato@issp.u-tokyo.ac.jp}
\affiliation{Institute for Solid State Physics, the University of Tokyo, Kashiwa, Chiba 277-8581, Japan}

\date{\today}% It is always \today, today,

\begin{abstract}
We present a theoretical study of the quantum critical behavior in heat transport via a two-state system with sub-ohmic reservoirs.
We calculate the temperature dependence of the thermal conductance near the quantum phase transition via the continuous-time quantum Monte Carlo method and discuss its critical exponents.
We also propose a superconducting circuit to realize the sub-ohmic spin-boson model that can be used to observe quantum critical phenomena.

\end{abstract}

\pacs{Valid PACS appear here}% PACS, the Physics and Astronomy
\maketitle

%\tableofcontents

\input{MainSec1}
\input{MainSec2}
\input{MainSec3}
\input{MainSec4}

\section*{Acknowledgement}
We also thank K. Saito for close discussions and critical reading of the manuscript.
The authors thank R. Sakano and T. Tamaya for helpful comments.
T.K. was supported by JSPS Grants-in-Aid for Scientific Research (No. JP24540316 and JP26220711).

\appendix

\input{AppCriticalExponent}
\input{AppCotunneling}
\input{AppCircuitModel}
\input{AppSpectralFunc}

\bibliography{Yamamoto_QPT}

\end{document}

%% file: MainSec1.tex
\section{Introduction}

% Its observation in mesoscopic device, e.g. two-channel Kondo

Quantum critical phenomena (QCP) induced by second-order quantum phase transitions (QPTs) are a central topic in condensed matter physics~\cite{Sachdev2011}.
Although QPTs have been studied in various highly-correlated systems, it is still challenging to realize them in controlled experimental systems.
Recently, QCP have been studied for the multi-channel Kondo effect realized in artificial nano structures~\cite{Potok2007,Mebrahtu2012,Mebrahtu2013,Keller2015,Iftikhar2015,Iftikhar2018}, and quantum critical behavior observed experimentally via electronic transport properties is in good agreement with theoretical results~\cite{Cox1998,Vojta2006,Bulla2008}.
This great success encourages further study of QCP in transport properties using different mesoscopic systems.

% General introduction: Quantum Phase transition
Heat transport in nano structures is another important topic in mesoscopic physics.
In particular, heat transport carried by photons (phonons) via a two-state system has been studied in several theoretical works~\cite{Segal2010,Ruokola2011,Saito2013,Ren2010,Chen2013,Segal2014,Yang2014,Wang2015,Taylor2015}, because it has considerable similarities to electronic transport in quantum dots.
The heat transport via a two-state is described by the spin-boson model, whose properties are characterized by the spectral density function $I(\omega)\propto\omega^s$~\cite{Leggett1987,Weiss2012}.
For sub-ohmic reservoirs ($0<s<1$), this model displays a QPT at zero temperature when a system-reservoir coupling is tuned to a critical value~\cite{Kehrein1995,Kehrein1996,Winter2009,Bulla2003,Vojta2005,Vojta2009,Vojta2012,Chin2011,Weiss2012}.
In a recent paper by the authors and the other two co-authors~\cite{Yamamoto2018}, the temperature dependence of thermal conductance is studied in detail for all types of reservoirs (arbitrary $s$) via continuous-time quantum Monte Carlo (CTQMC) simulations.
For sub-ohmic reservoirs, however, QCP near the transition point have not been discussed.

% Measurement
The recent, rapid development in nano structure fabrication and experimental heat measurement that has enabled us to experimentally access heat current in nano scale objects is remarkable~\cite{Forn-Diaz2017,Magazzu2017,Ronzani2018}. 
It has been demonstrated that transmission lines coupled to a superconducting qubit indeed realize the spin-boson model with an ohmic ($s=1$) reservoir~\cite{Yu2012,Bourassa2009,Leppakangas2018,Peropadre2013,Forn-Diaz2017,Magazzu2017}.
However, to the best of our knowledge, the design of a superconducting circuit to realize the sub-ohmic spin-boson model has only been discussed in Ref.~\cite{Tong2006}, in which experimental realization of the sub-ohmic reservoirs of $s=0.5$ is discussed.
To study QCP, considering the realization of the sub-ohmic spin-boson model for an arbitrary value of $s$ is advantageous.

% Heat transport
In this paper, we investigate QCP in heat transport via a two-state system carried by photons or phonons for sub-ohmic reservoirs.
The temperature dependence of the thermal conductance is calculated using the CTQMC method~\cite{Rieger1999,Winter2009,Yamamoto2018}.
In the previous work~\cite{Yamamoto2018}, it has been shown that the thermal conductance is always proportional to $T^{2s+1}$ at low temperatures when the system-reservoir coupling is below a critical value, reflecting a non-degenerate ground state of the system.
However, in the quantum critical regime near QPT, the power of the temperature dependence changes into a different value, reflecting the nature of QPT.
We discuss the critical exponents related to QPT in detail.
We also consider a superconducting circuit to realize the sub-ohmic spin-boson model with arbitrary value of $s$.

% Structures of sections
This paper is organized as follows.
The spin-boson model is described in Sec.~\ref{sec:Model}, and the heat current via a two-state system is formulated in Sec.~\ref{sec:Formulation}.
The critical temperature dependence of the heat current near the quantum phase transition is shown in Sec.~\ref{sec:result}, which is our main result.
A superconducting circuit is proposed that could be used to realize the spin-boson model with sub-ohmic reservoirs in Sec.~\ref{sec:realization}.
Finally, our results are summarized in Sec.~\ref{sec:summary}.
% Detailed discussions on the critical exponents and the co-tunneling formula are given in Appendix~\ref{app:CriticalExponent} and \ref{app:Cotunneling}, respectively.
%We also give detailed derivation on the spin-boson model from the circuit model in Appendix~\ref{app:CircuitModel}, and on the spectral function of the circuit model in Appendis~\ref{app:spect}, respectively.
Throughout this paper, we employ the unit of $k_{\rm B}=\hbar=1$.

%% file: MainSec2.tex
\section{Model}
\label{sec:Model}

\begin{figure}[tbp]
	\centering
	\includegraphics[width=7.0cm]{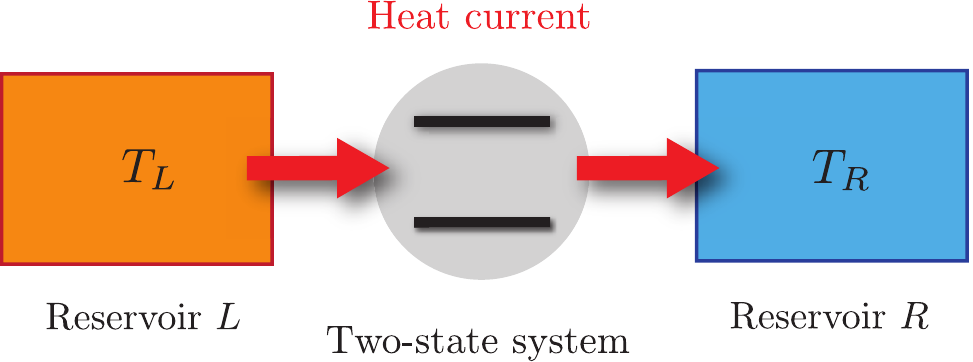}
	\caption{
    Schematic of the model comprises a two-state system coupled to two bosonic reservoirs ($L$ and $R$) with temperatures $T_{L}$ and $T_{R}$, respectively. If $T_L>T_R$, a heat current flows from reservoir $L$ to reservoir $R$ via the two-state system.
    }
	\label{FIG:model_transport}
\end{figure}

We consider heat transport between two bosonic reservoirs via a two-state system (see Fig.~\ref{FIG:model_transport}).
The model Hamiltonian is given by $H = H_{\rm S} + \sum_{\nu} H_{{\rm B},\nu} + \sum_{\nu} H_{{\rm I},\nu}$, where $H_{\rm S}$, $H_{{\rm B},\nu}$, and $H_{{\rm I},\nu}$ describe a two-state system, a bosonic reservoir $\nu$ ($=L,R$), and the system-reservoir coupling, respectively.
Each term of the Hamiltonian is given as follows:
\begin{eqnarray}
& & H_{\rm S} = -\frac{\Delta}{2}\sigma_x - \varepsilon\sigma_z,
\label{eq:H_system} \\
\label{eq:H_trans_B}
& & H_{{\rm B},\nu} = \sum_{k}\omega_{\nu k}b^\dagger_{\nu k}b_{\nu k}, \\
\label{eq:H_trans_I}
& & H_{{\rm I},\nu} = - \frac{\sigma_z}{2}\sum_{k}	\lambda_{\nu k}(b^\dagger_{\nu k} + b_{\nu k}),
\end{eqnarray} 
where $\sigma_\alpha$ ($\alpha=x,y,z$) is the Pauli matrix, and $b_{\nu k}$ ($b_{\nu k}^{\dagger}$) is an annihilation (a creation) operator of bosonic excitation with the wavenumber $k$ in the reservoir $\nu$.
The Hamiltonian of the two-state system, $H_{\rm S}$, is obtained by truncating a double-well potential system with the lowest two eigenstates, where $\Delta$ and $\varepsilon$ are the tunneling amplitude and detuning energy, respectively.
The energy dispersion of the reservoirs and the system-reservoir coupling strength are denoted by $\omega_{\nu k}$ and $\lambda_{\nu k}$, respectively.
In this paper, we consider heat transport for the symmetric case ($\varepsilon=0$). 
The detuning energy, $\varepsilon$, is used only for the detailed discussion on critical exponents in Appendix~\ref{app:CriticalExponent}.

The property of the reservoirs is determined by the spectral density function:
\begin{equation}
\label{eq:spectral_trans}
I_{\nu}(\omega) \equiv \sum_{k}\lambda_{\nu k}^2\delta(\omega-\omega_{\nu k}) .
\end{equation}
For simplicity, the spectral density function is taken in the following form:
\begin{eqnarray}
& & I_{\nu}(\omega)  = \alpha_{\nu}\tilde{I}(\omega), \\
	\label{eq:spectral_tilde}
& & \tilde{I}(\omega) = 2\omega_{\rm c}^{1-s}\omega^s e^{-\omega/\omega_{\rm c}},
\end{eqnarray}
where $\alpha_{\nu}$ is the dimensionless system-reservoir coupling strength, and $\omega_{\rm c}$ is the cutoff frequency, which results in much larger energies in comparison with other characteristic energies.
Herein, we focus on the sub-ohmic case ($0<s<1$), for which a second-order quantum phase transition occurs.

\section{Formulation}
\label{sec:Formulation}

The heat current operator from the reservoir $\nu$ into the two-state system is defined as follows:
\begin{eqnarray}
\label{eq:current_operator}
J_\nu &\equiv& -\frac{dH_{{\rm B},\nu}}{dt} = i[H_{{\rm B},\nu},H]\nonumber \\
&=& -i\frac{\sigma_z}{2} \sum_{k} \lambda_{\nu k} \omega_{\nu k} (-b_{\nu k}+b_{\nu k}^{\dagger}).
\end{eqnarray}
Using the standard procedure of the Keldysh formalism~\cite{Rammer1986,Jauho1994,Jauho2007}, the following Meir-Wingreen-Landauer-type exact formula~\cite{Meir1992} for the heat current is derived~\cite{Ojanen2008,Saito2013,Saito2008}:
\begin{eqnarray}
\label{eq:current}
\average{J_L} =\frac{\alpha \gamma_a}{8}\! \int_0^{\infty}\!\!\! d\omega \, \omega\, \mathrm{Im}[\chi(\omega)]\tilde{I}(\omega)\left[n_L(\omega)-n_R(\omega)\right],
\end{eqnarray}
where $\alpha=\alpha_L+\alpha_R$, $\gamma_a=4\alpha_L\alpha_R/\alpha^2$ is an asymmetric factor, $n_\nu(\omega)$ is the Bose-Einstein distribution in the reservoir $\nu$, and $\chi(\omega)$ is the dynamic susceptibility of the two-state system defined by
\begin{eqnarray}
\label{eq:dynamic_susceptibility}
\chi(\omega) = -i \int_0^{\infty} dt \, e^{i\omega t} \langle [\sigma_z(t),\sigma_z(0)] \rangle.
\end{eqnarray}
The thermal conductance is obtained from Eq. (\ref{eq:current}) as
\begin{eqnarray}
\label{eq:conductance}
\kappa &=& \lim_{\Delta T \rightarrow 0} \frac{\average{J_L}}{\Delta T} \nonumber \\
&=& \frac{\alpha\gamma_a}{8}\int_{0}^{\infty}d\omega~\mathrm{Im}
[\chi(\omega)]\tilde{I}(\omega)
\left[\frac{\beta\omega/2}{\mathrm{sinh}(\beta\omega/2)}\right]^2,
\end{eqnarray}
where $\Delta T = T_L - T_R$ and $\beta = 1/T$ ($=1/T_L=1/T_R$).
To evaluate the thermal conductance, the dynamic susceptibility, $\chi(\omega)$, must be calculated in thermal equilibrium.

We numerically calculate the dynamic susceptibility, $\chi(\omega)$, using CTQMC simulations (for details on the CTQMC method, refer to Refs.~\cite{Winter2009,Yamamoto2018}).
Using the CTQMC method, we calculate the spin-spin correlation function $C(\tau)=\Braket{\sigma_z(\tau)\sigma_z(0)}_{\rm eq}$, where $\sigma_z(\tau)$ is the imaginary time path ($0 < \tau < \beta$), and $\average{\cdots}_{\rm eq}$ indicates the thermal average.
The dynamic susceptibility is obtained as: 
\begin{eqnarray}
& & \tilde{C}(i\omega_n) = \int_0^\beta \! d\tau \, e^{i\omega_{n}\tau}C(\tau),  
\label{eq:analyticC1} \\
& & \chi(\omega) = \tilde{C}(i\omega_n\rightarrow\omega+i\delta).
\label{eq:analyticC2} 
\end{eqnarray}
The analytic continuation is performed numerically by the Pad\'e approximation~\cite{Baker1975,Vidberg1977}.

\section{Result}
\label{sec:result}

% Phase diagram, citing Winter,1 Yamamoto-preprint, very brief introduction on how to determine phase diagram (Binder parameter, etc.)

\begin{figure}[tbp]
	\centering
	\includegraphics[width=8.0cm]{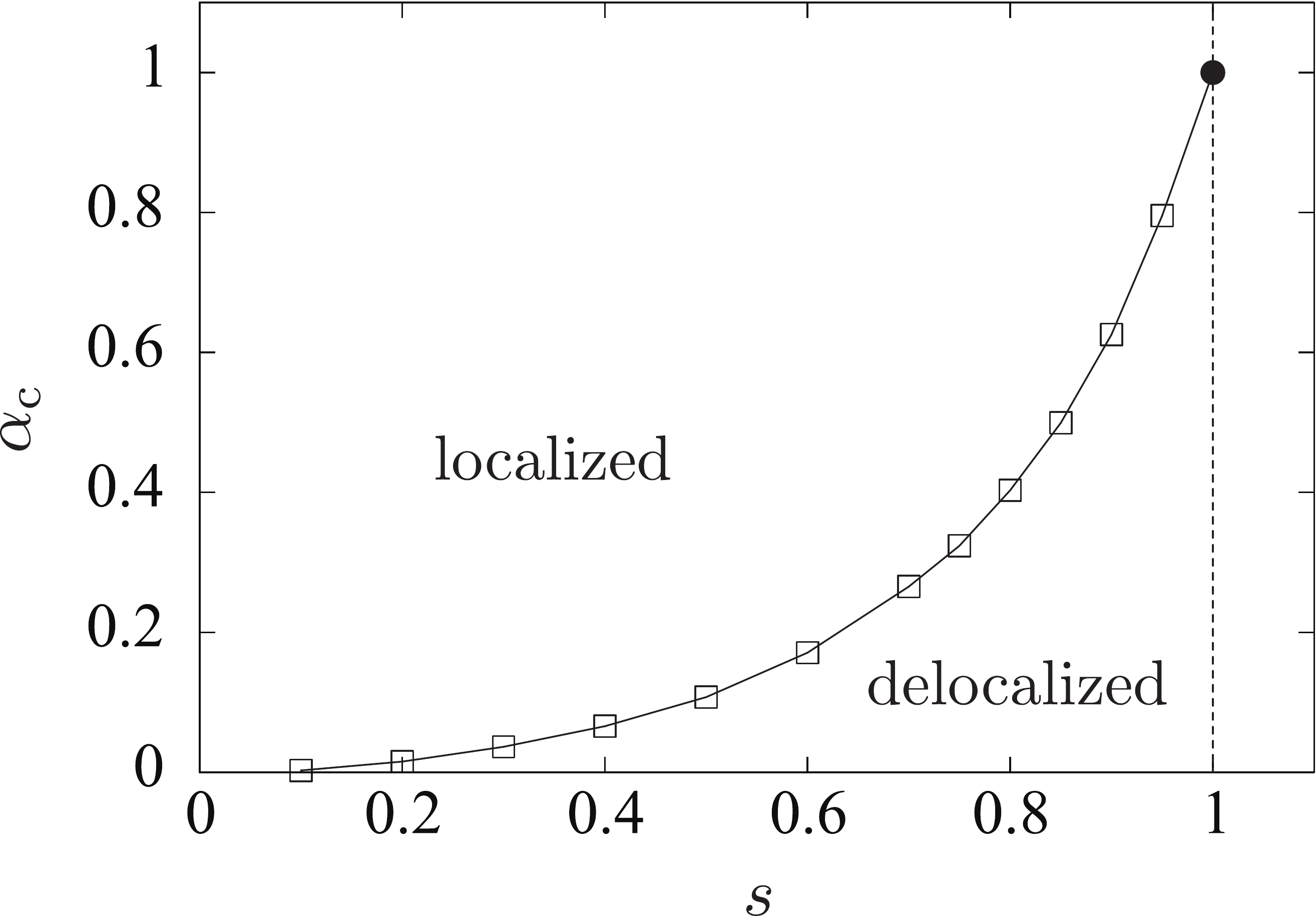}
	\caption{
	  The phase diagram of the sub-ohmic spin-boson model for $\Delta/\omega_{\rm c}=0.1$. 
      The solid line indicates the second-order transition line separating the delocalized and localized phases.
      The empty squares indicate the critical system-reservoir coupling that is numerically determined for the sub-ohmic case ($0<s<1$), whereas the filled circle represents the known transition point $\alpha_{\rm c}=1$ for the ohmic case ($s=1$).
}
\label{fig:phase_diagram}
\end{figure}

For the sub-ohmic case ($0<s<1$), a quantum phase transition occurs at zero temperature when the reservoir-system coupling reaches a critical value $\alpha_{\rm c}$, where $\alpha_{\rm c}$ is a function of $s$ and $\Delta/\omega_{\rm c}$~\cite{Kehrein1996,Bulla2003,Weiss2012}.
For $\alpha < \alpha_{\rm c}$, the ground state is described by a coherent superposition of two wave functions localized at each well ($\sigma_z = \pm 1$) and is called a ``delocalized state''.
For $\alpha > \alpha_{\rm c}$, the ground state becomes two-fold degenerate because the coherent superposition is completely broken owing to the disappearance of quantum tunneling between the two wells.
This state is called a ``localized state''.
The phase diagram of the spin-boson model determined by the CTQMC simulations for $\Delta/\omega_{\rm c}=0.1$ is shown in Fig.~\ref{fig:phase_diagram} (for details on determining the critical value, $\alpha_{\rm c}$, refer to Refs.~\cite{Volker1998, Winter2009,Yamamoto2018}).
The transition separating the two phases is of second-order for the sub-ohmic case (the empty squares) or of the Kosterlitz-Thouless-type~\cite{Chakravarty1982,Bray1982} for the ohmic case (the filled circle). 
This phase diagram is consistent with previous numerical studies~\cite{Winter2009,Bulla2003}.

\begin{figure}[tbp]
	\centering
	\includegraphics[width=8.0cm]{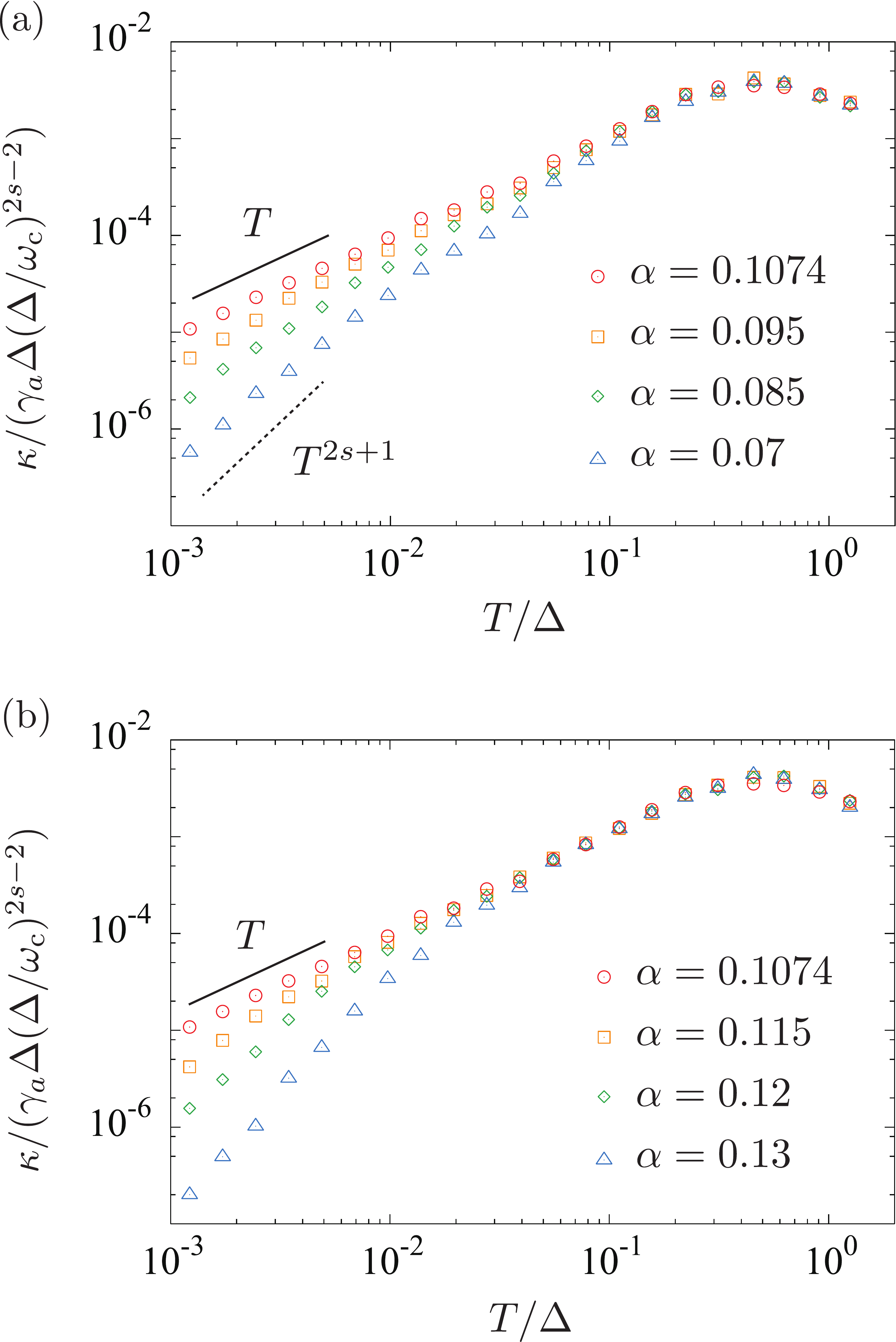}
	\caption{
	  The temperature dependence of the thermal conductance for (a) $\alpha \le \alpha_{\rm c}$ and (b) $\alpha\ge \alpha_{\rm c}$. 
     The plots represent the CTQMC simulation results for $s=0.5$ and $\Delta/\omega_{\rm c}=0.1$, for which the critical system-reservoir strength is $\alpha_{\rm c}=0.1074$.  }
	\label{fig:conductance}
\end{figure}

In Fig.~\ref{fig:conductance}, we show the temperature dependence of the thermal conductance for $s=0.5$ and $\Delta/\omega_{\rm c}=0.1$, where the critical system-reservoir coupling is $\alpha_{\rm c}=0.1074$.
Figs.~\ref{fig:conductance}~(a) and (b) show the delocalized-phase side ($\alpha \le \alpha_{\rm c}$) and the localized-phase side ($\alpha \ge \alpha_{\rm c}$), respectively.
%The horizontal axis is normalized by $\Delta$, whereas the vertical axis is normalized by a temperature-independent factor, $\gamma\Delta(\Delta/\omega_{\rm c})^{2s-2}$, with dimensions of the conductance.
In general, at the critical point, the thermal conductance exhibits distinctive power-law behavior determined by the nature of QPT:
\begin{equation}
\kappa \propto T^{c}, \quad (\alpha = \alpha_{\rm c}),
\label{eq:definitionc}
\end{equation}
where $c$ is the critical exponent dependent on $s$.
As shown in Fig.~\ref{fig:conductance}, the exponent $c$ is 1 for $s=0.5$.
As the system-reservoir coupling is reduced below the critical value ($\alpha < \alpha_{\rm c}$), the temperature dependence of the thermal conductance deviates from one at the critical point.
For a sufficiently small system-reservoir coupling (e.g., $\alpha=0.07$ in Fig.~\ref{fig:conductance}~(a)), the thermal conductance becomes proportional to $T^{2s+1}$ at low temperature, presumably for heat transport due to co-tunneling (see Appendix~\ref{app:Cotunneling}).
The temperature dependence of the thermal conductance also deviates as the system-reservoir coupling is increased above the critical value ($\alpha > \alpha_{\rm c}$).
Its temperature dependence cannot be explained by a simple formula such as the noninteracting-blip approximation, which is expected to hold in the localized phase~\cite{Yamamoto2018}, up to $\alpha = 0.13$.

Let us discuss the critical exponent, $c$, defined in Eq.~(\ref{eq:definitionc}) for general values of $s$.
The static susceptibility is expressed by:
\begin{eqnarray}
\chi_0 &=& \beta\average{\bar{m}^2}_{\rm eq}, 
\label{eq:chifluctuation}
\\
\label{eq:magnetization2}
\bar{m} &=& \frac{1}{\beta}\int_0^\beta d\tau~\sigma_z(\tau).
\end{eqnarray}
Combining Eq.~(\ref{eq:chifluctuation}) with Eq.~(\ref{eq:magnetization2}), the static susceptibility is expressed as $\chi_0 = \int_0^{\beta} d\tau C(\tau)$ with the spin-spin correlation function $C(\tau)=\langle \sigma_z(\tau) \sigma_z(0) \rangle_{\rm eq}$.
At the critical point, the spin-spin correlation function exhibits the power-law decay:
\begin{eqnarray}
C(\tau) = C(\beta-\tau) \sim \tau^{-\eta}, \quad (\omega_c^{-1} \ll \tau \ll \beta/2),
\end{eqnarray}
where $\eta$ is the critical exponent related to the spin dynamics.
Then, the temperature dependence of the static susceptibility at the critical point is obtained:
\begin{eqnarray}
\chi_0 \sim \beta^{1-\eta}.
\end{eqnarray}
By using Eqs.~(\ref{eq:analyticC1}) and (\ref{eq:analyticC2}), the critical behavior of the imaginary part of the dynamic susceptibility is obtained:
\begin{eqnarray}
\label{eq:dynamic_sus_behavior2}
{\rm Im}[\chi(\omega)] &\sim& \omega^{\eta - 1}.
\end{eqnarray}
Substituting this into Eq.~($\ref{eq:conductance}$), the thermal conductance at the critical point behaves as $\kappa \sim T^{c}$, where the exponent is given by:
\begin{eqnarray}
c = s + \eta.
\end{eqnarray}
The critical exponent $\eta$ is a function of $s$ and has been analyzed in previous theoretical studies~\cite{Luijten_thesis,Winter2009}.
The phase transition for $0<s\le 1/2$ belongs to the mean-field universality class and leads to $\eta = 1/2$.
This conclusion is consistent with the critical exponent $c=1$ obtained by the CTQMC simulation for $s=1/2$ (see Fig. \ref{fig:conductance}).
For $1/2<s<1$, $\eta$ is a nontrivial function of $s$ and is evaluated by the $\varepsilon$-expansion~\cite{Luijten_thesis} (see Appendix~\ref{app:CriticalExponent}).
In summary, the exponent of the thermal conductance is given as follows:
\begin{eqnarray}
c = \left\{ \begin{array}{ll}
s + 1/2 & (s \le 1/2), \\
1 - \varepsilon/2 - \varepsilon^2A(s)/3s + \mathcal{O}(\varepsilon^3) & (s > 1/2),
\end{array} \right.\\ \nonumber
\end{eqnarray} 
where $\varepsilon = 2s-1$ and $A(s) = s[\psi(1)-2\psi(s/2)+\psi(s)]$.

Finally, we emphasize that the critical behavior near QPT can be observed for other physical quantities~\cite{Luijten_thesis,Tong2006,Winter2009}. % More references should be referred.
We summarize the critical exponents for measurable quantities in Appendix~\ref{app:CriticalExponent}.

%% file: MainSec3.tex
\section{Experimental Realization}
\label{sec:realization}

\begin{figure}[tbp]
	\centering
	\includegraphics[width=8.0cm]{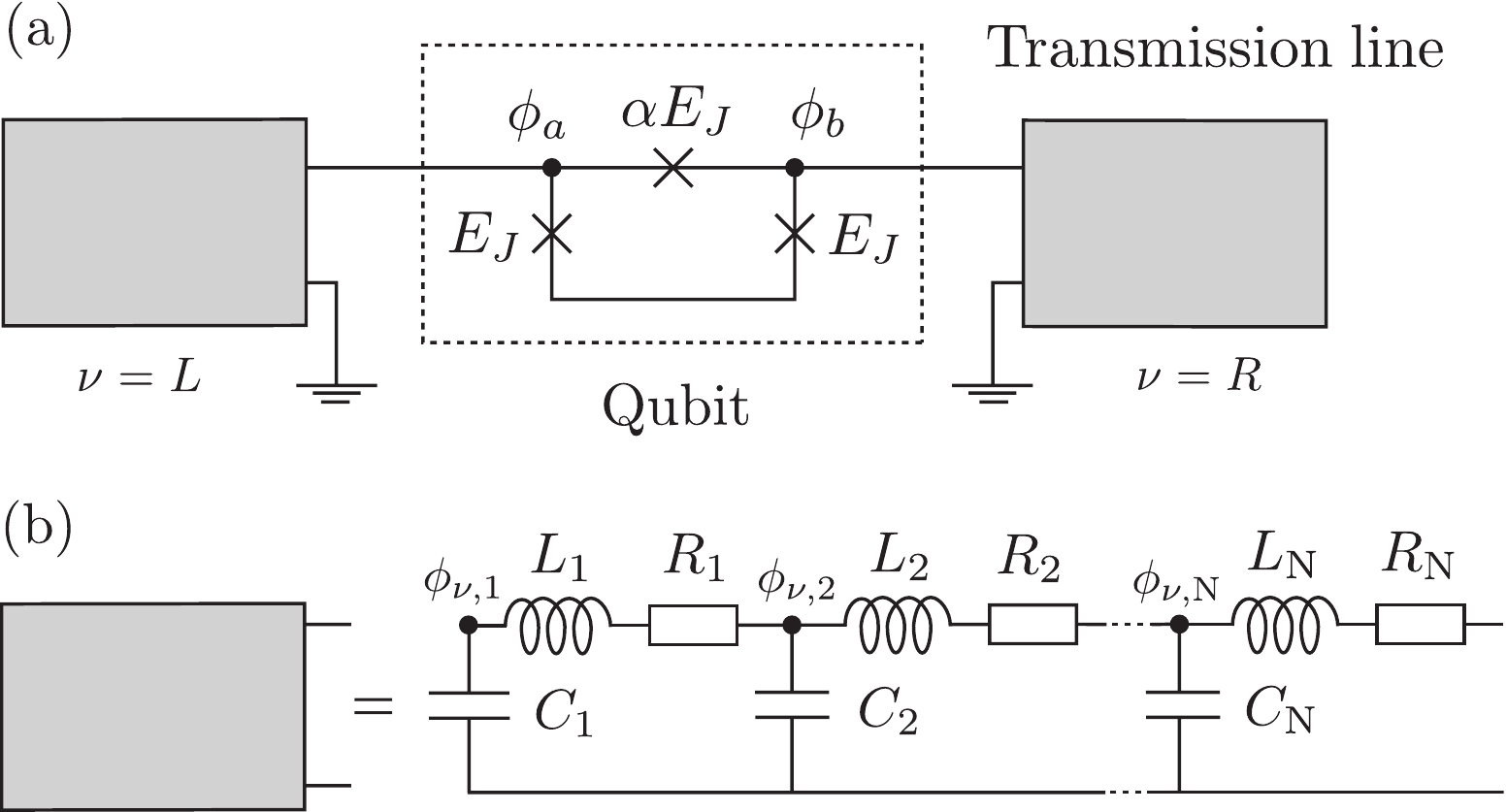}
	\caption{
	   	(a) A superconducting circuit composed of a flux qubit and two transmission lines.
        (b) The circuit of the transmission lines proposed to realize the sub-ohmic spin-boson model, consisting of resistances $R_{i}$, inductances $L_{i}$, and capacitances $C_{i}$.
    }
	\label{fig:model}
\end{figure}

In this section, we discuss a superconducting circuit that realizes a spin-boson Hamiltonian with sub-ohmic reservoirs.
A previous theoretical study~\cite{Tong2006} has shown that a spatially-uniform transmission line can realize a sub-ohmic reservoir with $s=0.5$.
For a controlled experiment of the QPT, however, it is favorable to realize a sub-ohmic reservoir with an arbitrary value of $s$.
We propose a superconducting circuit to realize a sub-ohmic reservoirs for arbitrary $s$ by introducing spatial dependence to the circuit elements.

We consider a flux qubit coupled to two transmission lines (or two junction arrays), as shown in Fig.~\ref{fig:model}~(a).
The flux qubit is composed of three small Josephson junctions~\cite{Mooij1999}.
By tuning the external magnetic field, the flux qubit acts like a double-well potential system, and its effective Hamiltonian is given by Eq.~(\ref{eq:H_system})
(for detailed derivation, see Appendix~\ref{app:CircuitModel}).
Then, the flux qubit coupled to the transmission lines can be described by the spin-boson model.
Using linear response theory~\cite{Schon1990,Tong2006,Bruus2004}, the spectral density function is expressed by the joint impedance of the two transmission lines ($Z(\omega) = \sum_{\nu} Z_\nu(\omega)$) as follows:
\begin{eqnarray}
	\label{eq:spectral_impedance}
    I(\omega) &=& \sum_{\nu} I_\nu(\omega) = \frac{4\phi_0^2\Braket{\varphi_-}^2}{\pi} I_0(\omega), \\
    I_0(\omega) &=& \omega{\rm Re}[Z(\omega)^{-1}],
    \label{eq:spectral_impedance2}
\end{eqnarray}
where $\phi_0 = \hbar/2e$, and $\pm \Braket{\varphi_-}$ is an expectation value of the phase at the flux qubit.
Detailed discussion is given in Appendix~\ref{app:CircuitModel}.

To realize a sub-ohmic reservoir with an arbitrary exponent, $s$, we propose a superconducting circuit, as shown in Fig.~\ref{fig:model}~(b).
The circuit comprises resistances $R_j$, inductances $L_j$, and capacitances $C_j$ ($j=1,2,\cdots,N$).
For simplicity, we assume that the two transmission lines are constructed by the same circuit.
The joint impedance of the two transmission lines is then calculated as $Z(\omega) = 2 Z_N(\omega)$, where $Z_j(\omega)$ ($j = 1,2,\cdots, N$) is given by a recurrence relation:
\begin{eqnarray}
	\label{eq:recurrence}
	Z_j(\omega) = R_j + i\omega L_j + \frac{1}{Z_{j-1}(\omega)^{-1} + i\omega C_j},
\end{eqnarray}
with $Z_0(\omega)^{-1} = 0$.

\begin{figure}[tbp]
	\centering
	\includegraphics[width=8.0cm]{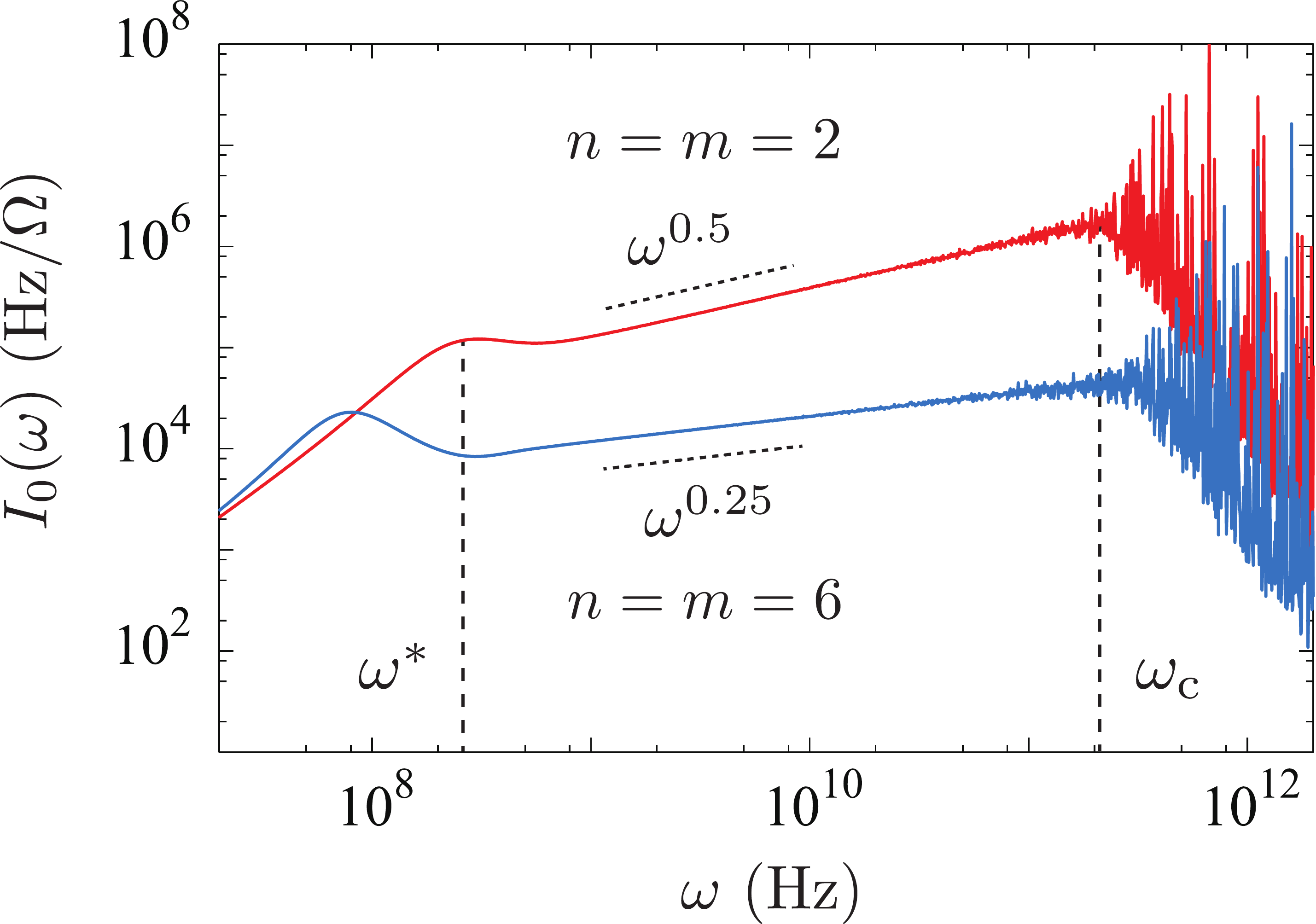}
	\caption{
	  The spectral density function of the superconducting circuit for $s=0.5$ and $0.25$, corresponding to $(n,m) = (2,2)$ and $(6,6)$.
      The circuit parameters are set as $N=10^4$, $R_0=1~{\rm k}\Omega$, $L_0=13~{\rm nH}$, and $C_0 = 1~{\rm pF}$.
    }
	\label{fig:spectral}
\end{figure}

Now, we assume that circuit elements have spatial dependence:
\begin{eqnarray}
	\label{eq:setting}
	& & R_j = R_0(1-j/N)^n, \\
   & & L_j = L_0, \\
   & & C_j = C_0(1-j/N)^m,
\end{eqnarray}
where $n$ and $m$ are non-negative real numbers.
We show the spectral density function, $I_0(\omega)$, of this circuit in Fig.~\ref{fig:spectral} for $(n,m) = (2,2)$ and $(6,6)$.
The parameters are set to $R_0 = 1\, {\rm k}\Omega$, $L_0 = 13 \, {\rm nH}$, $C_0 = 1 \, {\rm pF}$, and $N = 10^4$ and referred to experimental studies on Josephson junction arrays~\cite{Miyazaki2002}.
In Fig.~\ref{fig:spectral}, we added $1\%$ relative randomness for each circuit element to introduce tolerance to circuit parameter fluctuations.

We determined that the spectral density function is approximately proportional to $\omega^s$ in a certain range of the frequency with the exponent $0<s<1$.
This indicates that the present circuit can realize a sub-ohmic reservoir with an arbitrary value of $s$.
Certainly, the analytic calculation concludes: 
\begin{eqnarray}
	I(\omega) \propto \omega^{2/(m+2)}, \quad
    (\omega^* \ll \omega \ll \omega_{\rm c}).
\end{eqnarray}
The detailed calculation is given in Appendix~\ref{app:spect}.
This result is in good agreement with Fig.~\ref{fig:spectral}; $m=2$ and $6$ corresponds to $s=0.5$ and $0.25$, respectively.
The lower frequency limit for the sub-ohmic spectral density function, $\omega^*$, is calculated as follows:
\begin{eqnarray}
	\omega^* = \left[\left(\frac{m}{2N}\right)^{2n}\frac{R_0^{m+2}}{C_0^{n}L_0^{m+n+2}}\right]^{1/(m+2n+2)}.
\label{eq:omegastar}
\end{eqnarray}
Therefore, the exponent $n$ for the resistance~(\ref{eq:setting}) controls the lower limit of the sub-ohmic spectral density function.
In contrast, the higher frequency limit, $\omega_{\rm c}$, is a complex function of the circuit parameters.

In summary, the conditions for realizing a quantum phase transition are as follows:
First, the tunneling amplitude, $\Delta$, must be in the range of $\omega^* \ll \Delta \ll \omega_{\rm c}$.
Second, the dimensionless system-reservoir coupling, $\alpha$, should be tuned around the predicted critical point, $\alpha_{\rm c}$. 
For a typical value of the tunneling amplitude, $\Delta = 25~{\rm GHz}$, for the flux qubit~\cite{Magazzu2017}, we determined that both of the conditions are satisfied for the parameters used in Fig.~\ref{fig:spectral} for $s = 0.5$ ($m = 2$).
For this parameter set, the critical behavior of the thermal conductance at QPT described by Eq.~(\ref{eq:definitionc}) is expected in the temperature range of $\omega^* < T < \Delta$, when the system-reservoir coupling is tuned as $\alpha_{\rm c}$. 

%% file: MainSec4.tex
\section{Summary}
\label{sec:summary}

We studied quantum critical phenomena in heat transport by using a spin-boson model with sub-ohmic reservoirs.
By implementing continuous-time quantum Monte Carlo simulations, we show that the thermal conductance at the critical point has a characteristic power-law temperature dependence determined by the nature of QPT.
We also clarify the means by which the critical exponent of the thermal conductance is related to other critical exponents discussed in previous theoretical studies.
Finally, we propose a superconducting circuit that realizes sub-ohmic reservoirs for an arbitrary value of the exponent $s$.

We expect that our study will provide a new platform for experiments attempting to access quantum phase transitions directly upon measuring the transport properties of mesoscopic devices.
Although we used the flux qubit to realize the spin-boson model, other types of qubits such as a charge qubit or a transmon qubit could be considered.
We will present detailed descriptions of the other types of qubits in other studies.

%% file: AppCriticalExponent.tex
\section{Critical Exponents}
\label{app:CriticalExponent}

\begin{figure}[tbp]
	\centering
	\includegraphics[width=8.0cm]{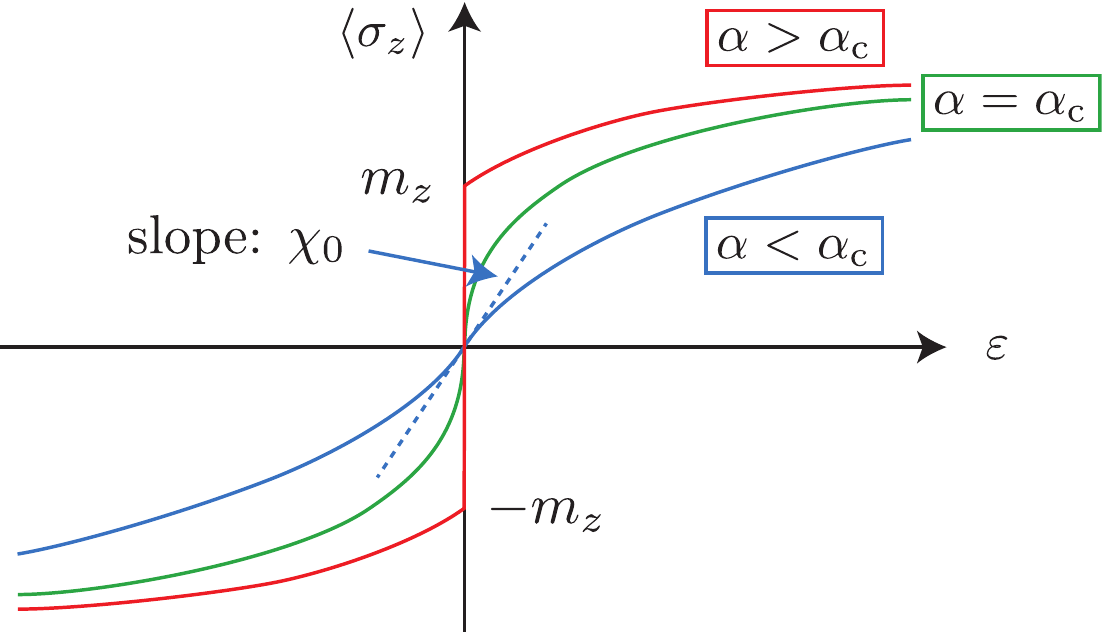}
	\caption{
Population, $\Braket{\sigma_z}$, as a function of the detuning energy, $\varepsilon$. For the delocalized phase (blue line; $\alpha<\alpha_{\rm c}$), $\Braket{\sigma_z}$ is a continuous function of $\varepsilon$, and the susceptibility, $\chi_0$, can be defined by the slope at $\varepsilon=0$. At the critical point (green line; $\alpha=\alpha_{\rm c}$), $\Braket{\sigma_z}$ is continuous, but the susceptibility diverges at $\varepsilon=0$. For the localized phase (red line; $\alpha>\alpha_{\rm c}$), $\Braket{\sigma_z}$ is discontinuous at $\varepsilon=0$.
}
\label{fig:QPT}
\end{figure}

% Behavior of $\langle \sigma_z \rangle$ using figure.

In this Appendix, we briefly discuss the critical exponents of several observables at the quantum phase transition for sub-ohmic reservoirs~\cite{Bulla2003,Vojta2005,Winter2009}.
Fig.~\ref{fig:QPT} shows schematics of the population, $\average{\sigma_z}$, as a function of the detuning energy, $\varepsilon$, near the critical point, $\alpha=\alpha_{\rm c}$.
In the delocalized phase ($\alpha < \alpha_{\rm c}$), the slope at $\varepsilon = 0$ corresponds to the static susceptibility:
\begin{eqnarray}
	\label{eq:static_susceptibility}
	\chi_0 = \lim_{\varepsilon\rightarrow0}\frac{\average{\sigma_z}_{\rm eq}}{\varepsilon}.
\end{eqnarray}
The static susceptibility, $\chi_0$, diverges as the value of $\alpha$ approaches $\alpha_{\rm c}$ from below.
In the localized phase ($\alpha > \alpha_{\rm c}$), $\average{\sigma_z}$ jumps from $-m_z$ to $m_z$ at $\varepsilon = 0$, where $m_z= \average{\sigma_z}|_{\varepsilon\rightarrow +0}$ is the spontaneous magnetization.

\begin{table}[tb]
  \label{table:exponent}
  \caption{Summary of the critical exponents.}
  \begin{center}
    \begin{tabular}{lll} \hline \hline
      Exponent & Definition & Condition   \\ \hline
      $\gamma $ & $\chi_0 \propto (\alpha_{\rm c}-\alpha)^{-\gamma}$ & $\alpha<\alpha_{\rm c}$, $T=0$ \\
      $\beta'$ & $m_z \propto (\alpha-\alpha_{\rm c})^{\beta'}$ & $\alpha>\alpha_{\rm c}$, $T=0$ \\
      $\eta$ & $m_z \propto T^{\eta/2}$ & $\alpha=\alpha_{\rm c}$, $T> 0$ \\
      $x$ & $\chi_0 \propto T^{-x}$ & $\alpha=\alpha_{\rm c}$, $T> 0$ \\ \hline \hline
    \end{tabular}
  \end{center}
\end{table}

In Table~\ref{table:exponent}, we summarize the critical exponents. 
All of the exponents can be determined experimentally by measuring the population, $\average{\sigma_z}$.
By using $y_h^*$ and $y_t^*$, two exponents related to the QPT fix point,
these critical exponents are expressed as follows~\cite{Winter2009}:
\begin{eqnarray}
& & \beta' = (1-y_h^*)/y_t^*, \\
& & \gamma = (2y_h^*-1)/y_t^*, \\
& & \eta = 1-x = 2-2y_h^*.
\end{eqnarray}
Since the transition occurs above the upper critical dimension,
for $0<s \le 0.5$, the exponents $y_t^*$ and $y_h^*$ are given by mean-field theory as follows:
\begin{eqnarray}
y_t^* = 1/2, \quad y_h^* = 3/4.
\end{eqnarray}
Therefore, we obtain 
\begin{eqnarray}
\beta' = 1/2, \quad \gamma = 1, \quad \eta = 1/2, \quad x = 1/2.
\end{eqnarray}
For $s > 0.5$, $y_t^*$ and $y_h^*$ are nontrivial functions of $s$.
By the $\varepsilon$-expansion, the exponents are calculated as~\cite{Luijten_thesis}:
\begin{eqnarray}
\label{eq:y_t}
& & y_t^* =s + \varepsilon/6 - 4\varepsilon^2A(s)/9s + \mathcal{O}(\varepsilon^3),\\
\label{eq:y_h}
& & y_h^* = (1+s)/2 + \varepsilon/4 - \varepsilon^2A(s)/6s + \mathcal{O}(\varepsilon^3),
\end{eqnarray}
where $\varepsilon = 2s-1$, $A(s) = s[\psi(1)-2\psi(s/2)+\psi(s)]$, and $\psi(x)$ is the Digamma function.
The results of the critical exponents are confirmed in previous numerical studies~\cite{Luijten_thesis,Luijten1997,Winter2009,Vojta2005}.

%% file: AppCotunneling.tex
\section{Asymptotically-Exact Formula for Co-tunneling}
\label{app:Cotunneling}

When the ground state is a delocalized state ($\alpha < \alpha_{\rm c}$), heat transport is induced by the virtual excitation of the two-state system for $T \ll \Delta_{\rm eff}$, where $\Delta_{\rm eff}$ is a renormalized tunneling amplitude.
This process is called co-tunneling.
By utilizing the generalized Shiba relation~\cite{Sassetti1990}, the asymptotically-exact formula for the thermal conductance in the co-tunneling regime ($T \ll \Delta_{\rm eff}$) is derived as follows~\cite{Yamamoto2018}:
\begin{eqnarray}
\label{eq:conductance_co}
\kappa_{\rm co} = \frac{\pi\chi_0^2}{8}\int_0^\infty d\omega~I_L(\omega)I_R(\omega)\left[\frac{\beta\omega/2}{\sinh(\beta\omega/2)}\right]^2,
\end{eqnarray}
where $\chi_0$ is the static susceptibility defined by Eq.~(\ref{eq:static_susceptibility}).
This formula leads to thermal conductance proportional to $T^{2s+1}$.

%% file: AppCircuitModel.tex
\section{Circuit Model}
\label{app:CircuitModel}

In this appendix, we consider a flux qubit coupled to transmission lines (see Fig.~\ref{fig:model}) and derive the effective spin-boson model~\cite{Peropadre2013}.
For example, we consider a uniform transmission line with constant capacitance and inductance  ($C_i = C$, $L_i = L$) while neglecting-resistance ($R_i = 0$).
We then derive a general linear response relation between the spectral density function and the circuit impedance.

The Hamiltonian of the present circuit is given by
\begin{eqnarray}
	\label{eq:H_circuit_dis}
    & & H = H_{\rm S} + H_{\rm B} + H_{{\rm I}}, 
\\
    \label{eq:H_qubit1}
    & & H_{\rm S} = \sum_{k=1}^{3} \left[\frac{Q_{J,k}^2}{2C_{J,k}} - E_{J,k}\cos(\phi_{J,k}/\phi_{0})\right],\\
    \label{eq:H_transmission_dis}
    & & H_{{\rm B}} = \sum_{\nu}  \sum_{j=1}^{N} \left[\frac{Q_{\nu,j}^2}{2C} + \frac{(\phi_{\nu,j+1}-\phi_{\nu,j})^2}{2L}\right], \\
    \label{eq:H_int_dis}
    & & H_{{\rm I}} =    
    \frac{(\phi_{a}-\phi_{L,N})^2}{2L_N}
 + \frac{(\phi_{R,N}-\phi_{b})^2}{2L_N},
\end{eqnarray}
where $H_S$, $H_{{\rm B}}(=\sum_\nu H_{{\rm B},\nu})$, and $H_{\rm I}(=\sum_\nu H_{{\rm I},\nu})$ describe the flux qubit, the transmission lines, and the system-reservoir coupling, respectively, and $\phi_0 = \hbar/2e$ is the flux quantum.
The flux qubit comprises three Josephson junctions with Josephson energies $E_{J,k}$ ($k=1,2,3$), and the charge and flux operator of the $k$-th Josephson junction are denoted by $Q_{J,k}$ and $\phi_{J,k}$, respectively.
Similarly, the charge and flux operators of the transmission line (see Fig.~\ref{fig:model}~(b)) are denoted by $Q_{\nu,j}$ and $\phi_{\nu,k}$, respectively, and these operators satisfy the exchange relations $[\phi_{J,k},Q_{J,k'}] = i\delta_{k,k'}$ and $[\phi_{\nu,j},Q_{\nu',j'}] = i\delta_{j,j'}\delta_{\nu,\nu'}$, respectively.
The flux operators at the two sides of the flux qubit are expressed by $\phi_a$ and $\phi_b$ (refer Fig.~\ref{fig:model}~(a)). 

To make the flux qubit, the area of one junction is reduced by a factor of $\alpha$ ($E_{J,1} = E_{J,3} = E_J$, $C_{J,1} = C_{J,3} = C_J$, $E_{J,2} = \alpha E_{J}$, and $C_{J,2} = \alpha^{-1} C_{J}$).
Then, the Hamiltonian of the flux qubit Hamiltonian~(\ref{eq:H_qubit1}) can be rewritten~\cite{Mooij1999,Peropadre2013}
\begin{eqnarray}
	\label{eq:H_qubit2}
    H_{\rm qb} &=& \frac{Q_{J,+}^2}{2C_{J,+}} + \frac{Q_{J,-}^2}{2C_{J,-}} + V(\phi_{J,+},\phi_{J,-}), \\
    V(\phi_{J,+},\phi_{J,-}) &=& - E_{J}[2\cos(\phi_{J,+}/2\phi_{0})\cos(\phi_{J,-}/2\phi_{0}) \nonumber \\
    & & \hspace{5mm} + \alpha\cos((\Phi_{\rm ext}-\phi_{J,-})/2\phi_{0})],
\end{eqnarray} % Add the definition of C_+ and C_-
where $\phi_{J,\pm} = (\phi_{J,1} \pm \phi_{J,3})/2$, its conjugate operator is denoted by $Q_{J,\pm}$, and $V(\phi_{J,+},\phi_{J,-})$ is the Josephson energy that plays the role of the potential energy.
When the magnetic flux through the loop is tuned to be half of the flux quantum ($\Phi_{\rm ext}=\phi_0/2$), the Josephson energy, $V(\phi_{J,+},\phi_{J,-})$, has two energy minima on the line $\phi_{J,+}=0$.
Due to quantum tunneling effects, there is an energy splitting $\Delta$ between the ground state and the first-excited state.
Since these lowest two eigenstates are well separated from the other eigenstates, we can truncate the system into the lowest two eigenstates, thus leading to the two-state system Hamiltonian (\ref{eq:H_system}).
The wavefunctions of the lowest two states are described as $\ket{\sigma_x=+1} = (\ket{\uparrow} + \ket{\downarrow})/\sqrt{2}$ and
 $\ket{\sigma_x=-1} = (\ket{\uparrow} - \ket{\downarrow})/\sqrt{2}$, where $\ket{\uparrow}$ and $\ket{\downarrow}$ are the two-dimensional wavefunctions localized at the two potential energy minima, respectively

Introducing the new variables $\phi_{\pm} = \phi_{R,{\rm N}} \pm \phi_{L,{\rm N}}$ and $\Phi_{\pm} = \phi_{b} \pm \phi_{a}$ and using $\phi_{J,+} \propto \Phi_{+} \simeq 0$, the system-reservoir coupling~(\ref{eq:H_int_dis}) is rewritten as $H_{\rm I}  = -\phi_{-}\Phi_{-}/2L_N$.
After truncation into the two-state system, we obtain:
\begin{eqnarray}
	\label{eq:H_int_dis2}
	H_{\rm I} = -\frac{\phi_{-}}{2L_N}\phi_0\Braket{\varphi_{-}}\sigma_z,
\end{eqnarray}
where $\bra{\uparrow}\Phi_{-}\ket{\uparrow} \equiv \phi_0\Braket{\varphi_{-}}$, $\bra{\downarrow}\Phi_{-}\ket{\downarrow} \equiv -\phi_0\Braket{\varphi_{-}}$, and $\bra{\uparrow}\Phi_{-}\ket{\downarrow} = \bra{\downarrow}\Phi_{-}\ket{\uparrow} = 0$.

For simplicity, we consider the continuous limit $\Delta x \rightarrow 0$ while keeping the length of the transmission line, $L_t =N \Delta x$, constant, where $\Delta x$ is the size of each elementary island.
Then, the system-reservoir coupling can be rewritten by~\cite{Peropadre2013}:
\begin{eqnarray}
	\label{eq:H_int_con}
    & & H_{{\rm I}} = -\frac{1}{l}\left.\frac{\partial \phi(x)}{\partial x}\right|_{x = 0}\phi_0\Braket{\varphi_{-}}\sigma_z,
\end{eqnarray}
where $l$ is the inductance per unit length.
The flux, $\phi(x)$, can be expressed by:
\begin{eqnarray}
	\phi(x) = \sum_k \frac{1}{\sqrt{2c\omega_k}}(b_k+b_k^\dagger)\frac{e^{ikx}}{\sqrt{L_t}},
\end{eqnarray}
where $c$ is the capacitance per unit length, and $b_k$ and $b_k^\dagger$ are bosonic annihilation and creation operators, respectively.
Then, the Hamiltonians for the transmission lines and the system-reservoir coupling can be rewritten as follows:
\begin{eqnarray}
	\label{eq:H_transs}
    & & H_{{\rm B}} = \sum_{k} \omega_{k}b^\dagger_{k}b_{k}, \\
	\label{eq:H_int_con2}
	& & H_{{\rm I}} = -\frac{\sigma_z}{2}\sum_{k}\lambda_{k}(b_k+b_k^\dagger), \\
   & & \lambda_{k} = \frac{2\phi_0\Braket{\varphi_{-}}}{v l\sqrt{L_t}}\sqrt{\frac{\omega_k}{2c}},
   \label{eq:H_int_con3}
\end{eqnarray}
where $v = 1/\sqrt{lc}$ is the speed of light in the transmission line.
This model corresponds to the spin-boson model with an ohmic reservoir.

Now, we discuss the general linear response relation.
The electric current operator at the position $x$ is defined by $\mathcal{I}(x) = l^{-1} \partial \phi(x)/\partial x$ and is calculated at $x = 0$:
\begin{eqnarray}
	\label{eq:electric_current}
    \mathcal{I}_0 \equiv \mathcal{I}(x=0) = \sum_{k}\frac{i\lambda_{k}}{2\phi_0\Braket{\varphi_{-}}}(b_k+b_k^\dagger).
\end{eqnarray}
From Eqs.~(\ref{eq:spectral_trans}) and (\ref{eq:H_int_con2})-(\ref{eq:electric_current}), the spectral density function can be rewritten as: 
\begin{eqnarray}
	\label{eq:spectral_circuit}
    I(\omega) = \frac{4\phi_0^2\Braket{\varphi_{-}}^2}{\pi}{\rm Im}[G_{\mathcal{I}_0}^{\rm R}(\omega)],
\end{eqnarray}
where $G_{\mathcal{I}_0}^{\rm R}(\omega)$ is the Fourier transform of the current-current correlation function defined by $G_{\mathcal{I}_0}^{\rm R}(t) = -i\theta(t)\Braket{[\mathcal{I}_0(t),\mathcal{I}_0(0)]}$.
Using linear response theory~\cite{Bruus2004}, $G_{\mathcal{I}_0}^{\rm R}(\omega)$ can be related to the total impedance of the transmission lines:
\begin{eqnarray}
	\frac{1}{Z(\omega)} = \frac{i}{\omega} G_{\mathcal{I}_0}^{\rm R}(\omega).
    \label{eq:linearresponse}
\end{eqnarray}
Substituting Eq.~(\ref{eq:linearresponse}) into Eq.~(\ref{eq:spectral_circuit}), we can derive Eqs.~(\ref{eq:spectral_impedance}) and (\ref{eq:spectral_impedance2}) in the main text.
Although we have derived them for a special case, i.e., the case of uniform transmission lines without damping, Eqs.~(\ref{eq:spectral_impedance}) and (\ref{eq:spectral_impedance2}) hold for arbitrary circuits of the transmission lines.

%% file: AppSpectralFunc.tex
\section{Analytic Expression of the Spectral Density Function}
\label{app:spect}

We analyze the frequency dependence of the spectral density function for the circuit model discussed in Sec.~\ref{sec:realization}.
Assuming $|\omega C_j Z_{j-1}(\omega)| \ll 1$, the following recurrence relation (\ref{eq:recurrence}) is given approximately:
\begin{eqnarray}
Z_j(\omega) \simeq R_j + i\omega L_j + Z_{j-1}(\omega)-i\omega C_j Z_{j-1}(\omega)^2.
\end{eqnarray}
In the continuous limit $N \rightarrow \infty$, this recurrence relation reduces to the differential equation: 
\begin{eqnarray}
	\frac{dZ(\omega,x)}{dx} = r(x) + i\omega l(x) - i\omega c(x)Z(\omega,x)^2,
\end{eqnarray} 
where $r(x)$, $l(x)$, and $c(x)$ ($0 \le x =j/N \le 1$) are the resistance, inductance, and capacitance per unit length, respectively.
From Eq.~(\ref{eq:setting}), they are given as
\begin{eqnarray}
	& & r(x) = r_0(1-x)^{n}, \\
   & & l(x) = l_0, \\
   & & c(x) = c_0(1-x)^{m},
\end{eqnarray}
where $r_0=R_0/\Delta x$, $l_0=L_0/\Delta x$, and $c_0=C_0/\Delta x$.
We note that $Z(\omega) = Z(\omega,x\rightarrow 1)$.
Since $\dot{Z}(\omega,x) = dZ(\omega,x)/dx$ and $r_0(1-x)^{n}$ are sufficiently small compared with other terms, we can neglect them and obtain:
\begin{eqnarray}
	Z_A(\omega,x) = \sqrt{\frac{l_0}{c_0}}(1-x)^{-m/2},
\end{eqnarray}
for
\begin{eqnarray}
	\label{eq:condition_1-x}
	1-x^* \equiv\left(\frac{n}{2\omega\sqrt{l_0c_0}}\right)^{2/(m+2)} \! \! \! \! \! \ll 1-x \ll \left(\frac{\omega l_0}{r_0}\right)^{1/n} \! \!.
\end{eqnarray}
In contrast, for $x \simeq 1$, we can neglect $r(x)$ and $c(x)$, and obtain the following:
\begin{eqnarray}
	Z_B(\omega,x) = i\omega l_0 x + A(\omega).
\end{eqnarray}
The constant of integration, $A(\omega)$, can be determined by the equation $Z_A(\omega,x^*) = Z_B(\omega,x^*)$.
Thus, we arrive at $Z(\omega)$ as follows:
\begin{eqnarray}
	Z(\omega) &\sim& Z_B(\omega,x \rightarrow 1) \nonumber \\
	&=& i\omega l_0(1-x^*) + \sqrt{\frac{l_0}{c_0}}(1-x^*)^{-m/2}.
\end{eqnarray}
From Eq. (\ref{eq:spectral_impedance}), we obtain the following spectral density function:
\begin{eqnarray}
	I(\omega) \propto \omega{\rm Re}[Z(\omega)^{-1}] \propto \omega^{2/(m+2)}.
\end{eqnarray}
This frequency dependence appears for $\omega^* \ll \omega \ll \omega_{\rm c}$, where the lower bound, $\omega^*$, is obtained by considering the condition (\ref{eq:condition_1-x}):
\begin{eqnarray}
	\omega^* = \left[\left(\frac{m}{2}\right)^{2n}\frac{r_0^{m+2}}{c_0^{n}l_0^{m+n+2}}\right]^{1/(m+2n+2)}.
\end{eqnarray}
This corresponds to Eq.~(\ref{eq:omegastar}) in the main text.